\documentclass[10pt]{iopart}
\input{pdfcolor}
\usepackage{amssymb,amsfonts,latexsym, soul}
\usepackage{fancyhdr}

\usepackage{dcolumn}
\usepackage{subfigure, psfrag}

\def\RCS$#1: #2 ${\expandafter\def\csname RCS#1\endcsname{#2}}

\RCS$Date: 2008-10-21 15:01:43 $
\RCS$Revision: 1.56 $
\RCS$Date: 2008-10-21 15:01:43 $
\RCS$Revision: 1.56 $
\usepackage[pdftex]{graphicx}

\usepackage{ifpdf}
\ifpdf
\usepackage[pdftex,plainpages=false,pdfpagelabels,hypertexnames=false,colorlinks=true,pdfstartview=FitV,linkcolor=blue,citecolor=blue,urlcolor=blue]{
hyperref}
\usepackage{thumbpdf}
\else

\usepackage{hyperref}
\fi

\begin{document}
\bibliographystyle{unsrt}
\pagestyle{plain}
\fancypagestyle{plain}

\title{Searching for gravitational-wave signals emitted by eccentric compact
binaries using  non-eccentric template bank: implications for ground based-detectors}

\author{T.Cokelaer, D. Pathak} 
\address{School of Physics and Astronomy, Cardiff University, Cardiff CF24
3AA, UK}
\ead{Thomas.Cokelaer@astro.cf.ac.uk}
\ead{Devanka.Pathak@astro.cf.ac.uk}

\begin{abstract}
Most of the inspiralling compact binaries are expected to be circularized by
the time their gravitational-wave signals enter the frequency band 
of ground-based detectors such as LIGO or VIRGO. However, it is not
excluded that some of these binaries might still possess a
significant eccentricity at a few tens of hertz. Despite this possibility,
current search pipelines -- based on matched filtering techniques -- consider 
only non-eccentric templates. The effect of such an approximation on the loss of signal-to-noise ratio (SNR) has been investigated by Martel and Poisson, \textit{Phys.\ Rev.\ D.\ } \textbf{60} 124008, in the context of initial LIGO detector. They ascertained that non-eccentric templates will be successful at detecting eccentric signals. We revisit their work by incorporating current and future ground-based detectors and precisely quantify the exact loss of SNR.

In order to be more faithful to an actual search, we maximized the SNR over a template bank, whose minimal match is set to 95\%. For initial LIGO detector, we claim that the initial eccentricity does not need to be taken into account in our searches for any system with total mass $M \in [2-45]M_\odot$ if $e_0 \lesssim 0.05$ because the loss of SNR (about 5\%) is consistent with the discreteness of the template bank. Similarly this statement is also true for systems with $M \in [6-35]M_\odot$ and $e_0 \lesssim 0.10$. However, by neglecting the eccentricity in our searches, significant loss of detection (larger than 10\%) may arise as soon as $e_0 \geq 0.05$ for neutron-star binaries. We also provide exhaustive results for VIRGO, advanced LIGO and Einstein Telescope detectors. It is worth noticing that for Einstein Telescope, neutron star binaries with $e_0\geq0.02$ lead to a 10\% loss of detection.
\end{abstract}

\maketitle
\section{Introduction}

Inspiralling compact binaries are one of the most promising source of gravitational-wave signals for interferometric ground-based detectors such as LIGO~\cite{LIGO} and VIRGO~\cite{VIRGO}. The phase and amplitude of the signal strongly depend on the masses of the two compact objects, the spin effects (neglected in this paper) and eccentricity. However, it is known that most of the inspiralling compact binaries would have circularized by the time they  enter the lower cut-off frequency of the detectors ($F_L$ hereafter, which is about $20~\rm{Hz}$ to $40\rm{Hz}$). Indeed, the eccentricity of an isolated compact binary evolving under the effect of gravitational radiation reaction is reduced by a factor of three when the semi-major axis is halved~\cite{Peters1964}. For instance, even though the orbital eccentricity of the Hulse and Taylor binary pulsar is $e_0=0.617$ (from the last measurements published in~\cite{HulseTaylor}), it would decrease to $10^{-6}$ by the time the orbital period of the binary reaches 0.02~s~\cite{KrolakSchutz1987}. Therefore, any isolated binary star formed by stellar evolution would have becomes circularized by the time its signal become measurable by ground-based detectors.

Nevertheless, in addition to the usual route to binary star formation, there exists a different
formation scenario in densely populated regions (e.g., in globular clusters) where binary-binary interactions can produce hierarchical triplets of black holes. The inner binaries of these triplets (undergoing Kozai oscillations) can merge under gravitational radiation reaction with large eccentricity~\cite{MillerHamilton2002}. Consequently, even at the lower cut-off frequency of ground-based detectors, there are compact binaries with significant eccentricity. As predicted in~\cite{Wen2003}, about 30\% of the binaries will posses eccentricities larger than $0.10$ when their emitted gravitational-wave signals reach a frequency of $10~\rm{Hz}$. There is also semi-analytical description of the final stages of mergers of black-hole neutron-star(BHNS) binaries showing that they may transit to inspiralling compact binaries in eccentric orbits~\cite{Davies2005}.

More recently, in~\cite{OLeary2008}, the authors study the case of stellar mass black holes (BHs) that segregate and form a steep density cusp around supermassive black holes in galactic nuclei. They found that BH binaries that form this way in galactic nuclei have an expected rate of coalescence events detectable by Advanced LIGO about $1-1000$ per year, depending on the initial mass function of stars in galactic nuclei and the mass of the massive BHs. The authors also claimed that such BH binaries have significant eccentricities as they enter the LIGO band ($90\%$ with $e> 0.9$), and hence they are distinguishable from other circularized binaries. They also show that mergers of eccentric binaries can be detected to larger distances and greater BH masses than circular mergers, up to ($\sim 700M_\odot$). 

Despite the fact that binaries may have non-negligible eccentricities, current searches for inspiralling compact binaries neglects this parameter. For instance, the matched filtering technique that is deployed to analyze scientific runs of LIGO data considers templates with circularized orbits only~\cite{S3S4Joint}. Yet, matched filtering is sub-optimal if the template is not a faithful representation of the signal that is searched for. In a previous work, Martel and Poisson~\cite{MartelPoisson1999} studied the loss of SNR when signals from eccentric binaries are filtered with templates based on circularized binaries. They performed this study in the context of initial LIGO. For instance, in the case of a $(1.4+1.4)M_\odot$ binary with an initial eccentricity of 0.05, they found that only a 2.4\% loss of SNR is expected. This result combined with the idea that eccentricities are small when a gravitational-wave signal enters a detector's band, led to searches where eccentricity is systematically neglected. 

In Ref.~\cite{MartelPoisson1999}, the authors considered a continuous space of templates to compute the loss of SNR. In this paper, we consider a more realistic situation by using a discrete grid of templates, which is identical to the one used in LIGO searches (where eccentricity is neglected). We also extend our target sources not only to binary neutron star (BNS) and low mass  binary black hole (BBH) but also to large mass BBH and BHNS. Our goal is to provide a precise description of the loss of SNR when waveforms from eccentric inspiralling compact binaries are searched with waveforms from circularized binaries. 

In Sec.~\ref{sec:model}, we describe the relevant features of the inspiraling compact binaries.
We also emphasize the differences between our waveforms and those obtained in~\cite{MartelPoisson1999}. 
In Sec.~\ref{sec:simulations}, we describe the simulation protocol and present exhaustive results showing the loss of SNR for different detectors, namely initial LIGO, advanced LIGO, Einstein Telescope and VIRGO detectors. 
Finally, in Sec.~\ref{sec:conclusion},  we discuss the implications of our results for ground-based detectors and future directions for this work.

\section{The eccentric model}\label{sec:model}
\subsection{eccentric waveform calculation}
The evolution of the orbital elements of an eccentric binary was first calculated by Peters and Mathews~\cite{Peters1964,PetersMathews1963}. The explicit expressions of the gravitational-wave signals were first 
derived in~\cite{Wahlquist}, in the context of spacecraft Doppler detection of gravitational waves. 
Although eccentric waveforms have been derived to a higher Post-Newtonian order~\cite{DamourGopakumarIyer}, we restrict ourselves to the case of Newtonian phase order. We use the same notation as in~\cite{MartelPoisson1999}. Let us start from the orbital radius $r$ that is given by 
\begin{equation}
r = \frac{pM}{1+e \cos\phi}\,,
\end{equation}
where $p$ is the semi-latus rectum, $M$ is the total mass, $e$ is the eccentricity, and $\phi$  is the orbital phase. The semi-major axis is related to $p$ and $e$ by the relation 
\begin{equation}\label{eq:a_functionof_p}
a = \frac{pM}{1- e^2}\,.
\end{equation}
Note that the orbital period is defined by 
\begin{equation}\label{eq:P}
 P=2\pi M \left(\frac{p}{1-e^2} \right)^{\frac{3}{2}}\,. 
\end{equation}

The decay of orbital elements $p$ and $e$ are calculated in the
quadrupole approximation~\cite{Peters1964}. Together with the orbital phase evolution, we have three ordinary differential equations (ODE) that can be used to describe the evolution of the system and form the starting point to determine the eccentric waveforms:
\begin{equation}\label{eq:dphidt}
\frac{d\phi}{dt}=\frac{\left(1+e \cos \phi \right)^2}{p^{\frac{3}{2}}M}\,,
\end{equation}

\begin{equation}\label{eq:dpdt}
 \frac{dp}{dt}=-\frac{64}{5}\frac{\mu}{M^2}\frac{\left( 1 -
e^2\right)^{3/2}}{p^3}\left(\ 1 + \frac{7}{8}e^2 \right) \,,
\end{equation}

\begin{equation}\label{eq:dedt}
 \frac{de}{dt} = -\frac{304}{15}\frac{\mu}{M^2}\frac{\left(1 - e^2 \right)^{3/2}
}{p^4} e \left( 1 + \frac{121}{304}e^2\right)\,.
\end{equation}

In these equations $\mu = m_1m_2/(m_1+m_2)$ is the reduced mass and,  $m_1$ and $m_2$ are the two component masses. We can convert Eqs.~(\ref{eq:dphidt}), (\ref{eq:dpdt}), and  (\ref{eq:dedt}) into the same notation as in~\cite{Peters1964}, by replacing $p$ by $a$ using Eq.~(\ref{eq:a_functionof_p}). 


In order to compute eccentric waveforms, we need to solve the
system of 3 ODEs as defined by Eqs~(\ref{eq:dphidt}), (\ref{eq:dpdt}), and (\ref{eq:dedt}). The method employed is described in details in
Sec.~\ref{sec:ode}.

Once the evolution of the three parameters $e$, $p$ and $\phi$ are known, we can generate the eccentric waveforms. 
The source and the gravitational wave detector are separated by a distance, $R$. The detector is placed in a direction
defined by the polar angles $\iota$ and $\beta$ relative to the Cartesian frame (following standard convention~\cite{Thorne}). The  unit vectors defined by $\hat{\iota}$ and $\hat{\beta}$ are chosen as polarization axes. The following equations are the two fundamental polarizations of the gravitational waves~\cite{Wahlquist}:
\begin{eqnarray}\label{eq:splus}
 s_{+} &=& -\frac{\mu}{pR} \Bigg\{  \bigg[ 2 \cos(2\phi - 2 \beta)  +  
 \frac{5e}{2}\cos(\phi - 2\beta) \nonumber \\ &+&
\frac{e}{2}\cos(3\phi-2\beta)+e^2\cos(2\beta) \bigg](1+\cos^{2}\iota)
 \nonumber \\ &+& [e\cos(\phi) + e^2]\sin^{2}\iota \Bigg\}. 
\end{eqnarray}
\begin{eqnarray}\label{eq:scross}
 s_{\times} &=& -\frac{\mu}{pR} \bigg[ 4\sin(2\phi-2\beta)+5e \sin(\phi -
2\beta)
\nonumber \\
&+& e\sin(3\phi-2\beta) - 2e^2\sin(2\beta) \bigg]\cos\iota.
\end{eqnarray}
In Sec.~\ref{sec:simulations}, we will see how $\iota$ and $\beta$ affect the waveform shape. It is important to notice that the signal can be decomposed into components that oscillate at once, twice and three times the orbital frequency. This effect is entirely related to the presence of eccentricity. Indeed if we set $e_0=0$, the remaining terms in Eqs~(\ref{eq:splus}) and (\ref{eq:scross}) contain components that oscillate at twice the orbital frequency only. 

We also need the initial conditions for the semi-latus rectum, $p_0$,  the orbital phase, $\phi_0$, and the eccentricity, $e_0$. The initial conditions on $e_0$ and $\phi_0$ are set arbitrarily and are specified in Sec.\ref{sec:simulations}. The remaining initial condition, $p_0$, is derived from the initial eccentricity, $e_0$. Using Eq.~(\ref{eq:P}), we have
\begin{equation}\label{eq:p0}
p_0= \frac{1-e_0^2}{\left(2\pi M f_0\right)^{2/3}}\,,
\end{equation}
where $f_0$ depends on the lower cut-off frequency, $F_L$, at which the signal enters a detector's sensitivity band. If we want the waveform to be valid from $F_L$ onwards, we must have the \textit{third} harmonic to start at $F_L$. The consequence is that the \textit{first} harmonic, which contributes to the third, must start at $F_L/3$. So, in Eq.~(\ref{eq:p0}), we must set $f_0=F_L/3$.

Finally, let us have a short digression about the effect of the parameter $f_0$, which may lead to difficulties from the point of view of data analysis. Indeed, if $e_0=0$ then the first and third harmonic contributions are null (see Eqs.~(\ref{eq:splus}) and (\ref{eq:scross})). Yet, with the  definitions provided above, we can see that the second harmonic starts at $2F_L/3$.  In the case of a non-eccentric waveform implementation, starting the waveform at $F_L$ is sufficient. So, if we compare the duration of an eccentric waveform (even when $e_0=0$) as defined in this paper  with the duration of a purely non-eccentric waveform then the ratio of the durations of two waveforms is $(2/3)^{-8/3}$. For	 instance, a $(1.4+1.4)M_\odot$ binary lasts 73 seconds if $F_L=40$~Hz, and 465 seconds if $F_L=20$~Hz. These durations have to be compared to a standard implementation where $f_0$ = $F_L$ : 25 and 158 seconds, respectively. Dealing with such long waveforms may lead to technical issues from a computational point of view (e.g., memory allocation).

\subsection{Eccentric waveforms: validation and investigations}\label{sec:ode}
The main difficulty in generating eccentric waveforms from Eqs.~(\ref{eq:splus}) and (\ref{eq:scross}) resides in the resolution of the system of differential equations. This is done by using a numerical code based on the GNU Scientific Library~\cite{GSL}. The ODE integrator is based on a Runge-Kutta-Fehlberg (4, 5) method. The step size is taken to be the inverse of the sampling frequency (4~kHz). Using this method, we compute the waveform durations as a function of initial eccentricity and component masses (see Table.~\ref{tab:duration}). Our results do not agree with those obtained in~\cite{MartelPoisson1999}: Durations generated by our code are consistently 5\% lower than those computed in~\cite{MartelPoisson1999}. This effect has been seen independently~\cite{vibha}. We performed another sanity check in the case where $e_0=0$: we compare our waveform with a non-eccentric waveform (at Newtonian order) that is generated with the LIGO Algorithm Library (LAL)~\cite{LAL}. We found a perfect agreement\footnote{To obtain the same waveform duration, we need to set the lower cut-off frequency of the non-eccentric waveform to $2 F_L/3$~Hz.} between the two waveforms. So, we think that the results in~\cite{MartelPoisson1999} may be biased due to a lack of accuracy in the evolution of the ODEs. We are now able to extend the waveform generation to higher masses and larger eccentricities as compared to Ref.~\cite{MartelPoisson1999}, whose study was limited to $(8.0+8.0)~M_\odot$.

The discrepancy that has been noted and the ability to extend investigations to higher mass range is also one of the motivation for revisiting their work.

In Table.~\ref{tab:duration}, we provide the duration of the eccentric waveforms, denoted $T(e_0, F_L)$, for various component masses and eccentricities.
Table~\ref{tab:duration} shows that for a given total mass, the duration of the signal decreases with increasing initial eccentricity. If a binary is highly
eccentric initially, then its orbit shrinks faster by losing energy at a faster
rate via radiation of gravitational waves. Moreover, for a given initial eccentricity,
the waveform duration decreases as the total mass increases. This behavior can be
anticipated by looking at Eq.~(\ref{eq:dpdt}), for which a larger initial value is expected for larger masses and larger eccentricity.

\begin{table}
\begin{indented}
\item[]
\begin{tabular}{c|llllll}

$e_0$ & 1.4+1.4     & 1.4+10       &  5+5        & 10+10    & 20+20   \\\hline
0     & 73.30(1.00)& 16.37(1.00) &  8.77(1.00) & 2.75(1.00) & 0.85(1.00) \\
0.1   & 70.66(0.96)& 15.78(0.96) &  8.46(0.96) & 2.65(0.96) &0.82(0.96)  \\
0.2   & 63.18(0.86)& 14.11(0.86) &  7.56(0.86) & 2.37(0.86) & 0.73(0.86) \\
0.3   & 52.09(0.71)& 11.63(0.71) &  6.23(0.71) & 1.95(0.71) &0.60(0.71)  \\
0.4   & 39.13(0.53)&  ~8.73(0.53) &  4.68(0.53) & 1.46(0.53) &0.43(0.50)  \\
0.5   & 26.22(0.36)&  ~5.84(0.36) &  3.13(0.36) & 0.97(0.35) &0.27(0.31)  \\
0.6   & 15.09(0.21)&  ~3.35(0.20) &  1.80(0.20) & 0.55(0.20) &0.06(0.07)  \\
0.7   &  ~6.92(0.09)&  ~1.52(0.09) &  0.82(0.09) & 0.23(0.08) &   /  \\
0.8   &  ~2.12(0.03)&  ~0.40(0.02) &  0.23(0.02) &   /        &   /  \\
\end{tabular}  
\end{indented}

\caption{\label{tab:duration}Duration of the gravitational-wave signal as a
function of initial eccentricity and component masses. The
lower cut-off frequency is $F_L=40$~Hz. The numbers provided in this table show
a systematic difference of about 5\% compared to numbers from  Tab.~I of~\cite{MartelPoisson1999}; we believe that our results are more accurate than
those from~\cite{MartelPoisson1999} (see the text for an explanation). Because
the duration decreases when eccentricity and total mass increase, there is a
combination for which the ending frequency of the waveform is below $F_L$. In such a case, no waveform can be generated, which is represented by the / symbol. In brackets, we also provide the ratio of the duration of an eccentric waveform with respect to the first line, where $e_0=0$. This ratio is independent of the component masses (See Fig.~\ref{fig:ratioTc} and the text for more details). } 
\end{table}

\begin{figure}[tbh]
\centering
\includegraphics[width=3.5in]{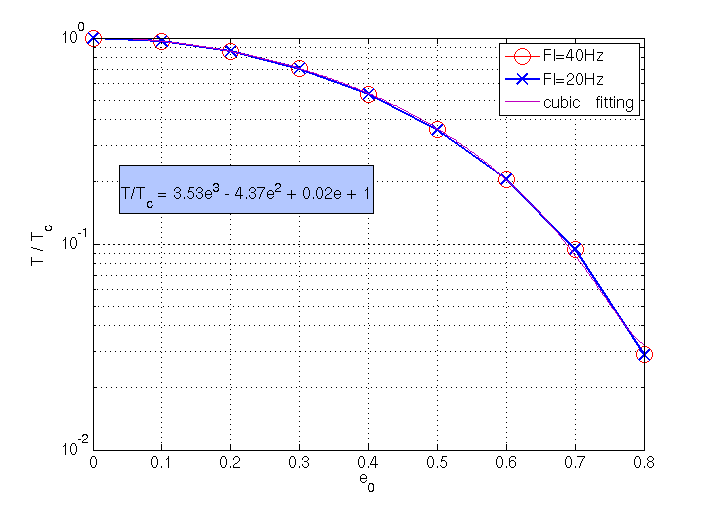}
\caption{\label{fig:ratioTc} The ratio of the duration of an eccentric binary
to that of a quasi-circularized waveform as a function of eccentricity. The durations of the eccentric ($T$) and circularized ($T_c$) systems are extracted from the integration of the ODEs. The ratio as a function of eccentricity can be accurately fitted with a cubic equation, which depends only on the initial eccentricity $e_0$. The crosses and 
circles show the ratio $T/T_c$ for two different lower cut-off frequencies $F_L=20$~Hz and $F_L=40$~Hz. The fit is also independent of $F_L$. It is also independent of the total mass (See Tab.~\ref{tab:duration}). }
\end{figure}

Using Tab.~\ref{tab:duration}, we can also compare the duration of the
circularized  waveform, $T_c(e_0=0)$, to the duration of the eccentric waveform, $T(e_0)$. The ratio of these two durations is shown in brackets. It is interesting to  notice that the ratio $T(e_0)/T_c(e_0=0)$ depends on the eccentricity only and is independent of the  total mass. Knowing the duration of the circularized waveform, we can anticipate the duration of the eccentric waveform by using the following expression 
\begin{equation}\label{eq:tbytc}
T(e_0) = T_c(e_0=0) \left[1+ 0.02 e_0 -4.37 e_0^2+3.53 e_0^3\right]\,.
\end{equation}
Although the derivation of Eq.~(\ref{eq:tbytc}) is empirical, it is accurate to 1\% error for eccentricity up to 0.6. This expression is independent of the initial cut-off frequency and masses of the system (in the Newtonian case) as shown in Fig.~\ref{fig:ratioTc}. This expression could be useful for data-analysis purposes so as to optimize  the vector's length to be stored in memory.

\begin{figure}[tbh]
\centering
\includegraphics[width=3.5in]{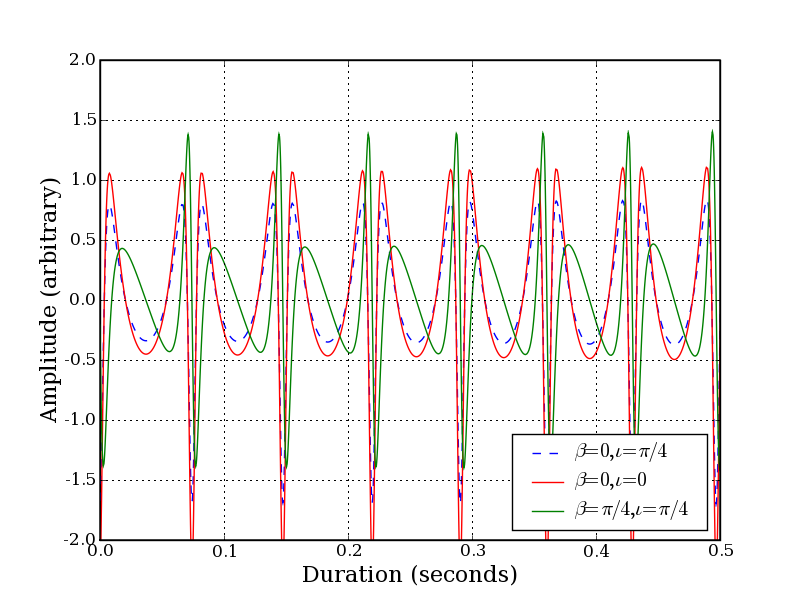}
\caption{\label{fig:waveform_beta_iota} Variation of a
$(5.0+5.0)M_\odot$ eccentric waveform with an initial eccentricity $e_0=0.5$ as a function of different $\beta$ and $\iota$ values.}
\end{figure}

Let us finish this section with an example. Let us consider a $(5.0+5.0)~M_\odot$  system with $e_0=0.5$. Duration of  such a waveform is $3.13~s$. We want to emphasize that the parameters $\beta$ and $\iota$ are of no importance for the simulations we present in Sec.~\ref{sec:search}. In Fig.~\ref{fig:waveform_beta_iota}, we plot the first $0.5~s$ of this waveform when $\beta=0$ and $\iota=\pi/4$. Then, we keep
$\beta=0$ and set $\iota=0$. We see that only the amplitude of the waveform
changes; the period remaining unchanged. Finally, we set
$\beta=\pi/4$ and $\iota=\pi/4$. Now, the amplitude changes as well as the phase.
However, the phase is only shifted. These results are expected if we look at Eq.~(\ref{eq:splus}) and Eq.~(\ref{eq:scross}), where we see that $\beta$ contributes to the phase, $\phi$, in the same manner in all the three harmonics, and $\iota$ comes as a factor for the amplitude only. So, neither $\beta$ nor $\iota$ affect the overall frequency behavior of the waveform. 

\section{Searching for eccentric binaries with templates from non-eccentric binaries}\label{sec:search}
In this section, we briefly describe the different tools that we have used in
our simulations. Then, we present the simulation protocol that we have followed. Finally, we present the results of our simulations in the context of different detectors and mass ranges. 

\subsection{Match and template bank}
We denote the Fourier transform of two functions $x(t)$ and $h(t)$ by
$\tilde{x}(f)$ and $\tilde{h}(f)$, respectively. The matched-filtering inner product is defined by
\begin{equation}\label{eq:mf}
\left(x , h\right) = 4 \int_{0}^{\infty} \frac{\tilde{x}^*(f) \tilde{h}(f) +
\tilde{x}(f) \tilde{h}^*(f)}{S_h(f)}df\,,
\end{equation}
where $S_h(f)$ is the noise power spectral density (PSD) of the detector. A normalized signal is given by
\begin{equation}
 \hat{h} = \frac{h}{\sqrt {\left(h, h\right)}}\,,
\end{equation}
where a hat denotes normalized signal. The SNR is defined by
\begin{equation}\label{eq:snr}
\rho(t) = \frac{\left(x, h\right)}{\sqrt {\left(h, h\right)}} = \left(x, \hat{h}
\right)\,.
\end{equation}

We can now imagine that $x(t)=n(t)+s(t,\vartheta^\mu)$, where $n(t)$ is Gaussian noise and $s(t,\vartheta^\mu)$ is a gravitational-wave signal characterized by a set of $p$ parameters $\vartheta^\mu\,, \mu=0,1,..,p-1$. These parameters may be separated into intrinsic and extrinsic parameters. The intrinsic parameters are represented by the component masses $m_1$ and $m_2$ and the initial eccentricity $e_0$, while the extrinsic parameters are the initial orbital phase $\phi_0$ and the time of arrival, $t_0$. When the noise fluctuations are neglected, we can define the match between a signal $s(t,\vartheta^\mu)$ and a template $h(t,\vartheta^\nu)$ as follows~\cite{OwenSathyaprakash98}:
\begin{equation}\label{eq:m}
 M\left(s(t,\vartheta^\mu),h(t,\vartheta^\nu\right))= \max_{t_0, \phi_0}\left(\hat{s}(t,\vartheta^\mu),
\hat{h}(t,\vartheta^\nu)\right)
\end{equation}
where the extrinsic parameters are maximized over automatically (using Fourier transform). 

Using a continuous template space, we can maximize the matches over the intrinsic parameters to obtain the fitting factor, $FF$, between the signal and the template family~\cite{Apostolatos1996}, which is defined by
\begin{equation}\label{eq:ff}
 FF\left(s(t,\vartheta^\mu\right),h(t,\vartheta^\nu))= \max_{\vartheta^\nu}M\left(\hat{s}(t,\vartheta^\mu),\hat{h}(t,\vartheta^\nu)\right)\,.
\end{equation}
The fitting factor can be interpreted as the maximum fraction of SNR that can be obtained by filtering a signal with an approximate template family. The signal and template families can be different, like in this paper, where the signal parameters are $\vartheta^\mu = m_1,m_2,e_0$ and the template parameters are $\vartheta^\nu = m_1,m_2$. The fitting factor was used in Ref.~\cite{MartelPoisson1999} to estimate the loss of SNR when an eccentric signal is filtered with a non-eccentric  signal.

In practice, the continuous template space is replaced by a discrete one, which is called a template bank~\cite{OwenSathyaprakash98, SD, Owen96}. It is represented by $\{h(t,\vartheta^\nu_i)\}$, where $i=0,1,..,N_b-1$, and $N_b$ is the number of templates. A template bank is optimally designed if $N_b$ is the smallest such that for any signal there always exists at least one template in the bank that gives
\begin{equation}\label{eq:mm}
 \min_{\vartheta^\mu}\max_{i} M\left(s(t,\vartheta^\mu),
h(t,\vartheta^\nu_i)\right) \geq
MM, 
\end{equation}
where $MM$ is the minimal match defined by the user. Usually, $MM$ is set to 95\%, or 97\%, which corresponds to a loss in detection rate of about 14\% and 9\%, respectively (the expected loss in detection equals $1 - MM^3$).
So, the minimal match is a measure of how well a discrete template bank covers the parameter space defined by the intrinsic parameters.

Let us define a new notation that will be useful to quantify our results. We define the match over the bank, $M_B$,   as follows
\begin{equation}\label{eq:mb}
 M_B(\hat{s}(t,\vartheta^\mu), {h(t,\vartheta^\nu_i}) =  \max_i M\left(\hat{s}(t,\vartheta^\mu),
\hat{h}(t,\vartheta^\nu_i)\right)\,.
\end{equation}
The quantity $1-M_B$ can be interpreted as the average fractional loss of SNR.

 Searches for inspiralling compact binaries performed in LIGO data use a template bank constructed so that signal from any circularized binary, $s(t)$, is found with a match greater than the minimal match, $MM$. Exhaustive simulations were performed in~\cite{squarebank,hexabank} to test the template bank placement in various cases: mass parameters related to BNS, BBH and BHNS systems and various design sensitivity curves of ground-based detectors. 

In the rest of this paper, we will study the distribution of the quantity $M_B$ when eccentric binaries are searched with the same template bank placement as the one used in LIGO analysis, where templates are generated with a non eccentric model.

\subsection{Simulation parameters}\label{sec:simulations}
As mentioned in the introduction, we want to identify the parameter space (in total mass, $M$, and initial eccentricity, $e_0$) where searching for eccentric binaries with templates corresponding to circularized systems suffices to obtain a negligible loss of SNR.
 
Our signals are based on eccentric waveforms whose phase is at Newtonian order only, we will  therefore also consider Newtonian order for our templates. We will use the so-called TaylorT3 approximant (see e.g,~\cite{hexabank, LAL} for a precise definition). The lower cut-off frequency of the template is chosen to be $F_L$. 

In all the simulations that follows, we use an hexagonal template bank with a minimal match $MM=95\%$. It has been shown that for VIRGO, LIGO, Einstein Telescope and  advanced LIGO design sensitivity curves,  $M_B$ is guaranteed to be larger than the minimal match (95\%) for any system with $ 2M_\odot \leq M \leq 60M_\odot$~\cite{hexabank}. Yet, in some cases, as described later in Sec.~\ref{sec:results}, we will extend the search to $80 M\odot$. Note that the template bank is actually optimized for 2PN order in phase and might therefore not be optimal for the Newtonian order considered here. In principle, since the phase of the signal and template are based on a Newtonian order only, we could use a one dimensional template bank (e.g., the chirp mass as defined later). So, using the template bank described in~\cite{hexabank} and used in LIGO searches, which is a two dimensional grid, our  simulations may provide slightly over estimated values of matches.

In all our simulations, we use the following common parameters. The number of waveforms in each simulation is 10,000. The sampling frequency is $4096$~Hz. The total mass is uniformly distributed in the range considered (see Sec.~\ref{sec:results}). The initial starting phase is randomized between 0 and $2\pi$. Although the choice of the initial phase should be of no consequence, we randomize it in all our simulations. The initial eccentricity is uniformly distributed in the range considered, that is $e_0 \in [0.0, 0.4]$. We also fix $\iota=\pi/4$, $\beta=0$. As explained in Sec.~\ref{sec:ode}, fixing $\beta$ and $\iota$ to arbitrary values does not affect the waveform significantly. However, we performed tests where $\beta$ and $\iota$ were randomized. As expected, we did not see any significant effects on the matches. Finally, let us note that if $\beta$ varies, then the ending frequency may slightly change and the matches as well. 

\begin{table}
\begin{indented}
\item[]
\begin{tabular}{c|cccc}
         & initial LIGO& Adv. LIGO & E.T. & VIRGO\\\hline
Bank size& 6792        &          25194 & 65104&52455\\
Bank size BNS & 2319 & 6969&16046 &12955 \\
\end{tabular}  
\caption{\label{tab:banksize}Number of templates, $N_b$, of each template bank used in our simulations. The template bank size of initial LIGO is the smallest because its lower cut-off frequency, $F_L$ equals 40~Hz whereas other detectors have $F_L=20$~Hz. The parameter space of the first row is $m_1,m_2 \in [1,30]M_\odot$ for initial LIGO and $m_1,m_2 \in [1,40]M_\odot$ for the others. The second row gives $N_b$ for a BNS parameter space, where $m_1,m_2 \in [1,3]M_\odot$ (see the appendix).}
\end{indented}
\end{table}

\subsection{Search for eccentric waveforms in ground-based detectors}\label{sec:results}
 We consider 4 design sensitivity curves: initial LIGO, advanced LIGO, Einstein Telescope and VIRGO. In this section we summarize the results we obtained when  searching for eccentric waveforms with circular waveforms.  We limited the maximum value of eccentricity to  $e_0=0.4$, above which the loss of SNR is always larger than 50\%. Note that for all the results presented below, the initial eccentricity is defined by Eq.~(\ref{eq:P}), which is defined at  $f_{\rm min}$; in other words, when the main harmonic frequency equals $2F_L/3$.

\subsubsection{Initial LIGO}

\begin{figure}[tbh]
\centering
\includegraphics[width=3.5in,height=2.5in]{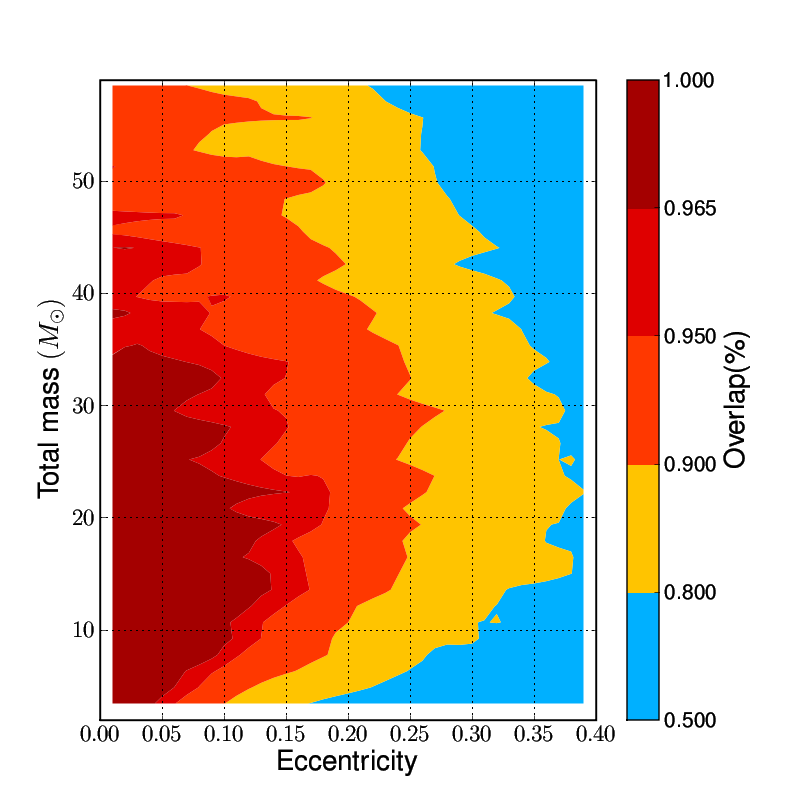}
\caption{\label{fig:ligoI}Maximum matches over the template bank ($M_B$) between an eccentric signal defined by its total mass (y-axis) and  initial eccentricity at $2F_L/3=26.66$~Hz (x-axis). The design sensitivity curve is based on initial LIGO. The colors represent the value of the maximum match over the template bank. The minimal match of the bank is 95\%. So the area where matches are below 95\% represent the region of parameter space where eccentricity significantly affect search for eccentric binaries with template based on circularized system.}
\end{figure}

In the case of initial LIGO, the lower cut-off frequency is $F_L=40$~Hz and the total mass range $M \in [2, 60]M_\odot$. The template bank size is provided in Tab.~\ref{tab:banksize}. In Fig.~\ref{fig:ligoI}, we show the distribution of the match over the bank, $M_B$, in the $(M, e_0)$ plane. As expected, around $e_0=0$, $M_B$ is close to 1. However, in most of the parameter space, $M_B$ is much smaller than unity. It is convenient to represent $M_B$ with isocontours set to [96.5, 95, 90, 80, 50]\%.  Using a value of 96.5\% is useful because the loss of detection associated with it is about 10\%. Using a value of 95\% is also useful because it corresponds to the minimal match of the template bank. 

The isocontours follow more or less the same structure. If we focus on the 95\% isocontour, we can see that BNSs are found with matches above 95\% if initial eccentricity is below 0.05. While the total mass increases up to $20~M_\odot$, the initial eccentricity for which the match is greater than 95\% increases up to a maximum value of 0.15. As the total mass increases for a match of 95\% or greater, from $20~M_\odot$ to about $40~M_\odot$, the initial eccentricity again goes down to 0.05. Above $40~M_\odot$, matches can go down to 90\% even for negligible eccentricities but this is related to the template bank design. Indeed, the template bank is optimally designed for signals that have a significant part of their power in a frequency band that resides where the detector has the best sensitivity. This assumption breaks down for signals with $M > 40M_\odot$, which ending frequency is already as low as 110~Hz whereas the detector best sensitivity is about 200~Hz. Consequently the bank is sub-optimal for these systems.

In the appendix, in Fig.~\ref{fig:chirpmass_match}, we provide for convenience the same results as in Fig.~\ref{fig:ligoI} where $M$ has been replaced by the chirp mass, $\mathcal{M}$, which is defined by $\mathcal{M}=M \eta^{3/5}$ and $\eta=\frac{m_1 m_2}{M^2}$. We also provide in Fig.~\ref{fig:lossdetection} the same results  as in Fig.~\ref{fig:ligoI} where $M_B$ has been replaced by the loss of detection ($1-M_B^3$).

Finally, since the parameter space related to BNSs is rather small in Fig~\ref{fig:ligoI}, we performed another simulation where we focus on the range $M \in [2, 6]M_\odot$. As we can see in Fig~\ref{fig:bns}, using non-eccentric waveforms is sufficient to detect a $(1.4+1.4)M_\odot$ eccentric waveforms if $e_0\leq0.05$.

\subsubsection{Advanced LIGO}
\begin{figure}[tbh]
\centering
\includegraphics[width=3.5in,height=2.5in]{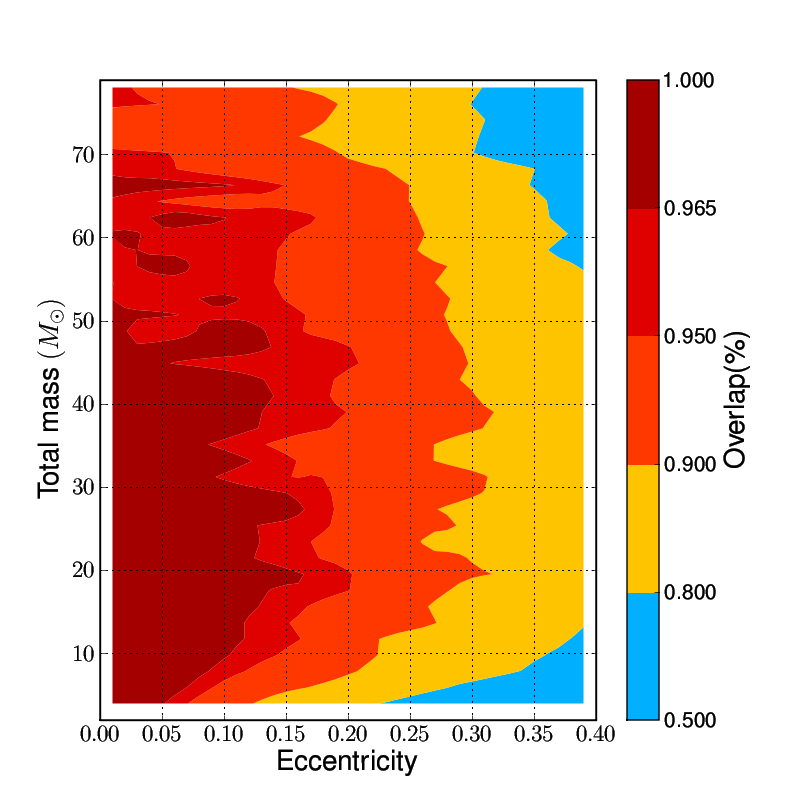}
\caption{\label{fig:ligoa} Same as Fig.~\ref{fig:ligoI} but for advanced LIGO.
}
\end{figure}

The simulation in the case of advanced LIGO differs from that of initial LIGO in two ways: the lower cutoff frequency is set to $F_L=20$~Hz, and the maximum total mass is extended to $80~M_\odot$. Results are shown in Fig.~\ref{fig:ligoa} and are very similar to the initial LIGO case. The main difference being that the high mass range is extended because the lower cut-off frequency is smaller.

If we focus on the 95\% isocontour, we can see that BNSs are found with matches above 95\% if initial eccentricity is below 0.05. Then, for the same match the initial eccentricity increases up to a maximum value of 0.15 as the total mass increases up to $25~M_\odot$. Finally, when the total mass increases from $25~M_\odot$ to about $70~M_\odot$, the  initial eccentricity remains constant around 0.15 for the same match of 95\%. Above $70~M_\odot$, matches go down to 90\% even for negligible eccentricities.

In the appendix (Fig.~\ref{fig:chirpmass_match}), we provide for convenience the same results as in Fig.~\ref{fig:ligoI} where $M$ has been replaced by the chirp mass. In Fig.~\ref{fig:lossdetection}), we also provide the same results  as in Fig.~\ref{fig:ligoI} where $M_B$ has been replaced by the loss of detection ($1-M_B^3$). We performed another simulation so as to focus on the BNS region. Results are shown in Fig.~\ref{fig:bns}:  using the non-eccentric model is sufficient to detect $(1.4+1.4)M_\odot$ eccentric systems with $e_0\leq0.05$.

\subsubsection{Einstein Telescope}
\begin{figure}[tbh]
\centering
\includegraphics[width=3.5in,height=2.5in]{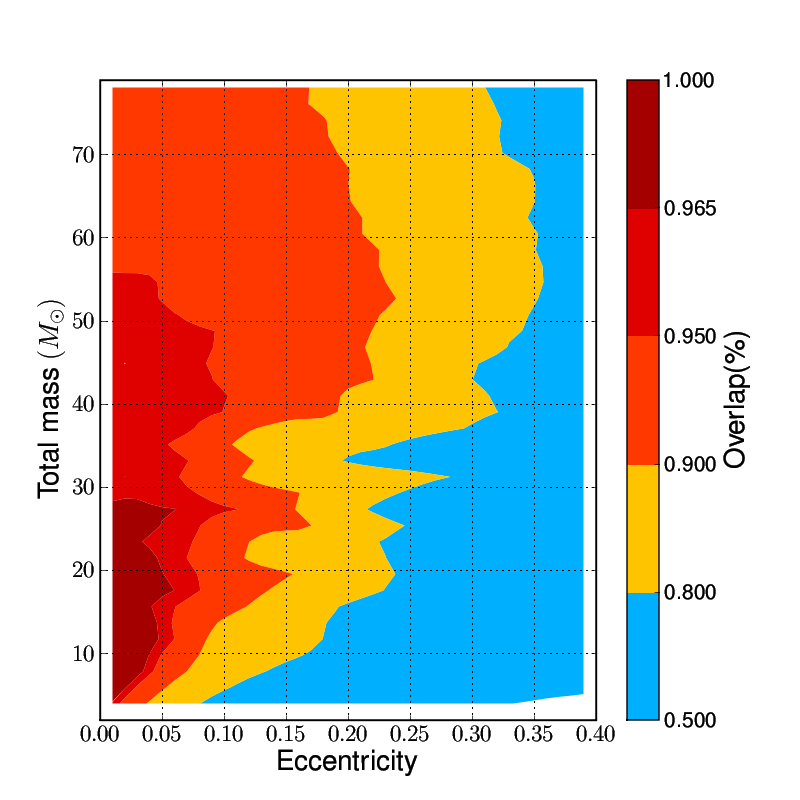}
\caption{\label{fig:ego} Same as Fig.~\ref{fig:ligoa} but for the Einstein Telescope.}
\end{figure}

The simulation in the case of the Einstein Telescope is similar to that of advanced LIGO (same $F_L$, same mass range). Results are shown in Fig.~\ref{fig:ego}. 

If we focus on the 95\% isocontour, we can see that BNSs are found with matches above 95\% if $e\sim0$. Then, the initial eccentricity increases up to a maximum value of 0.10 while total mass increases up to $30~M_\odot$ while obtaining a match $\geqslant 95\%$. Finally, when the total mass increases from $30~M_\odot$ to about $60~M\odot$, the initial eccentricity decreases to $e_0\sim0$. Among the four design sensitivity curves investigated in this paper, the Einstein Telescope case gives the least satisfactory results.

In the appendix, in Fig.~\ref{fig:chirpmass_match}, we replace $M$ by $\mathcal{M}$ using results from Fig.~\ref{fig:ego}. In  Fig.~\ref{fig:lossdetection}, we replace $M_B$ by the loss of detection ($1-M_B^3$). Finally, we focus on the area $M \in [2, 6]M_\odot$ and provide the results in Fig~\ref{fig:bns}: using non-eccentric model sufficient to detect a $(1.4+1.4)M_\odot$ eccentric system with $e_0\leq0.02$.

\subsubsection{VIRGO}
\begin{figure}[tbh]
\centering
\includegraphics[width=3.5in,height=2.5in]{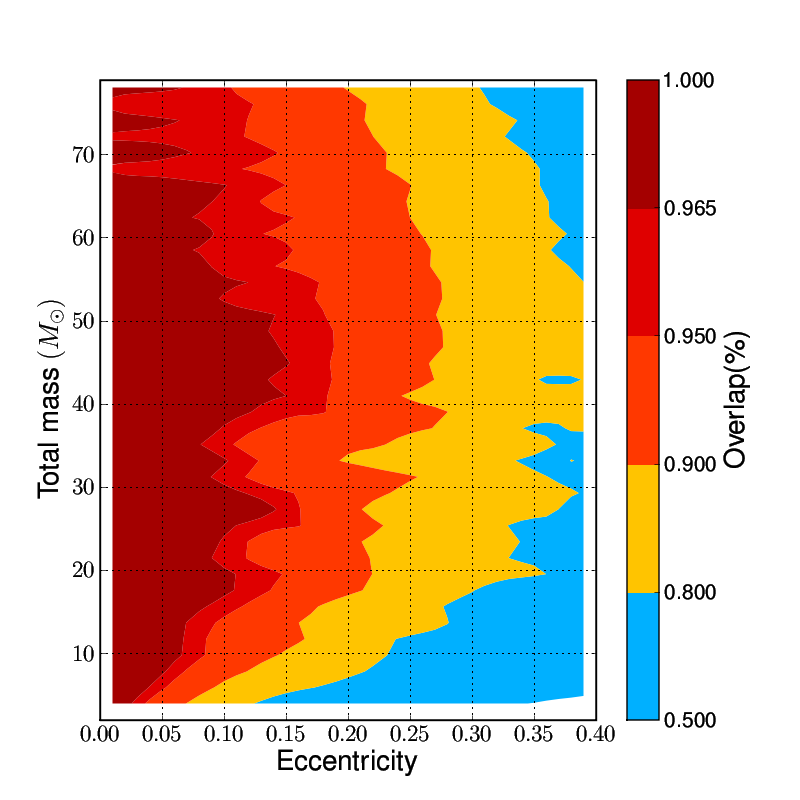}
\caption{\label{fig:virgo} Same as Fig.~\ref{fig:ligoa} but for VIRGO.}
\end{figure}

The simulation in the case of VIRGO is similar to that of advanced LIGO (same $F_L$, same mass range). The results are shown in Fig.~\ref{fig:virgo}. 

If we focus on the 95\% isocontour, we can see that BNSs are found with matches above 95\% if initial eccentricity is below 0.03. Then, the initial eccentricity can be increased up to maximum value of 0.15 while total mass increases up to $40~M_\odot$ which still achieve matches $\geqslant95\%$. Finally, when the total mass increases from $40~M_\odot$ to about $80~M\odot$, the  initial eccentricity decreases to $e_0=0.15$ for matches $\geqslant95\%$.

In the appendix, in  Fig.~\ref{fig:chirpmass_match}, we replace $M$ by $\mathcal{M}$ using results from Fig.~\ref{fig:virgo}. In Fig.~\ref{fig:lossdetection}, we replace $M_B$ by the loss of detection ($1-M_B^3$). Finally, we focus on the area $M \in [2, 6]M_\odot$ and provide the results in Fig~\ref{fig:bns}: using non-eccentric model is sufficient to detect a $(1.4+1.4)M_\odot$ eccentric binaries with $e_0\leq0.03$.

\section{Conclusion and perspectives}\label{sec:conclusion}
In this paper the main difference with previous works is that we maximize the matches over a discrete template bank, $M_B$, instead of computing the so-called fitting factor, $FF$, which maximizes the matches over a continuous parameter space. The template bank we used is identical to the template bank used in LIGO searches for binaries in quasi-circular orbits. For this reason, and since our estimation is more accurate, our results are slightly different from~\cite{MartelPoisson1999}.

If we search for gravitational-wave signal emitted by eccentric inspiralling compact binaries with standard non-eccentric template bank, our main results can be summarized as follows. In initial LIGO, (1) any binary neutron star system can be detected with $M_B \gtrsim 95\%$ if the initial eccentricity $e_0 \lesssim 0.05$ (or  $M_B \gtrsim 90\%$ if $e_0 \lesssim 0.10$) and (2) any binary system with a total mass, $ 6\leq M \leq 35~M_\odot$ can be detected with $M_B \gtrsim 95\%$ if $e_0 \lesssim 0.10$. Similar results have been obtained for the advanced LIGO case. Concerning VIRGO detector, any binary neutron star system can be detected with $M_B \gtrsim 95\%$ if $e_0 \lesssim 0.04$, or $M_B \gtrsim 90\%$ if $e_0 \lesssim 0.06$,  and (2) any system with total mass larger than $6~M_\odot$  can be detected with $M_B \gtrsim 95\%$ if $e_0 \lesssim 0.10$. Similar results have been obtained in the case of Einstein Telescope. In conclusion, the initial eccentricity does not need to be taken into account in our searches for gravitational-wave signals if $e_0 \lesssim 0.05$, because the loss of SNR (about 5\%) is consistent with the discreteness of the template bank that is being used.
However, by neglecting the eccentricity in our searches, significant loss of detection may arise as soon as $e_0 \geq 0.05$, especially for binary neutron stars and systems with high masses ($M \geq 35M_\odot$). 

Future directions for this work are the inclusion of higher PN order, the development of a template bank placement for eccentric systems and parameter estimation. Concerning higher PN order, we need to reproduce the simulations provided in this work. This work should be straightforward given the infrastructure which has been developed in this paper. Concerning the template bank placement, as we have explained, above $e_0=0.05$ the loss of SNR may be quite large in some region of the parameter space. A solution may be to use a dedicated template bank to take into account the eccentricity. For instance, using the template bank used in this paper, we can have several layers of templates, each having a different values of initial eccentricity. More investigations are needed but preliminary studies shows that only a few layers are required to significantly improve the results (the TaylorT3 family being replaced by the eccentric waveform presented here). Finally, using Monte-Carlo simulations it would be interesting to estimate the error bound on the variance of the measured eccentricity.

\ack{This research was supported partly by STFC, UK, grant PP/B500731. The authors benefited from useful discussions with members of the gravitational physics group at Cardiff University. In particular, the authors thank B.S~Sathyaprakash for suggesting this study.}

\appendix
\section{Additional results}

\begin{figure*}[]
\centering
\includegraphics[width=3in,height=1.8in]{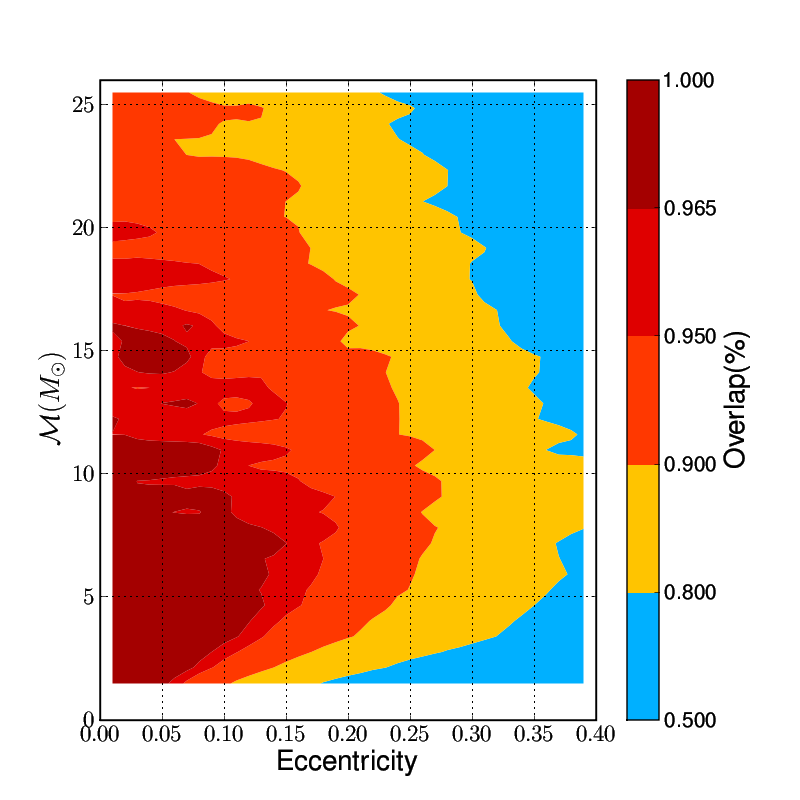}
\includegraphics[width=3in,height=1.8in]{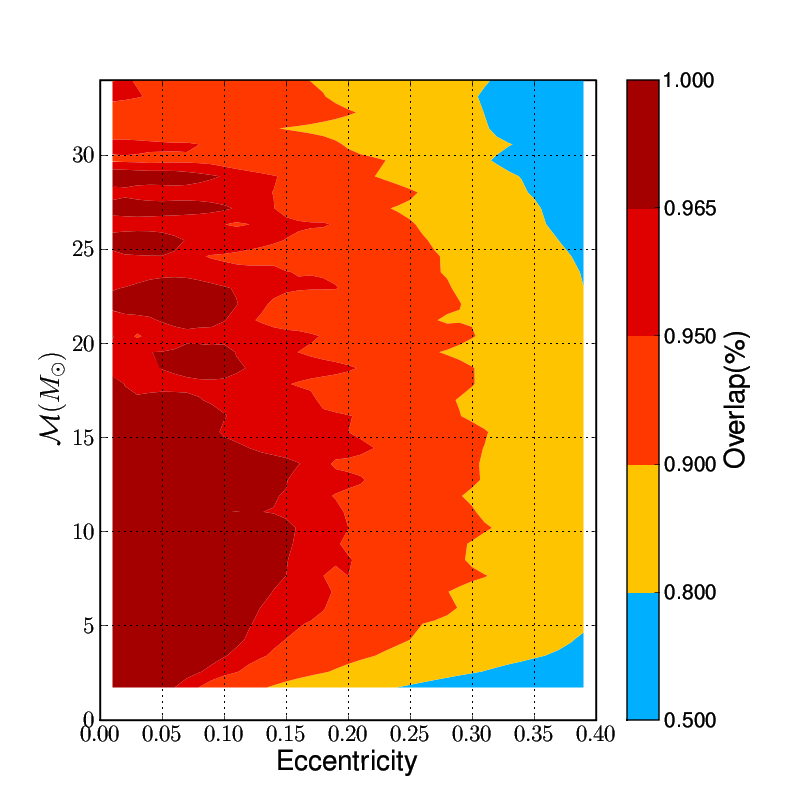}
\\
\includegraphics[width=3in,height=1.8in]{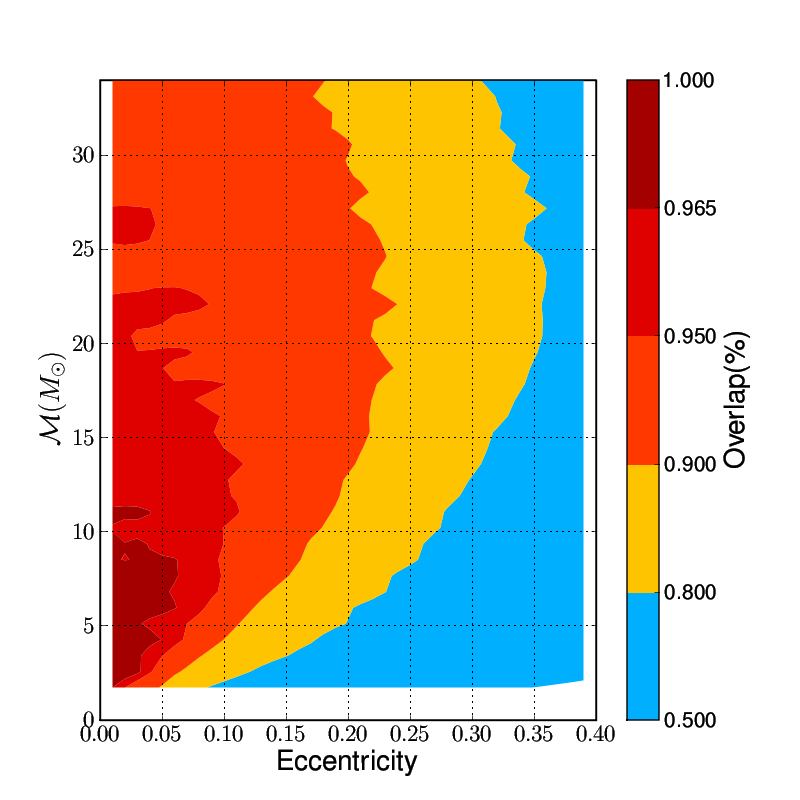}
\includegraphics[width=3in,height=1.8in]{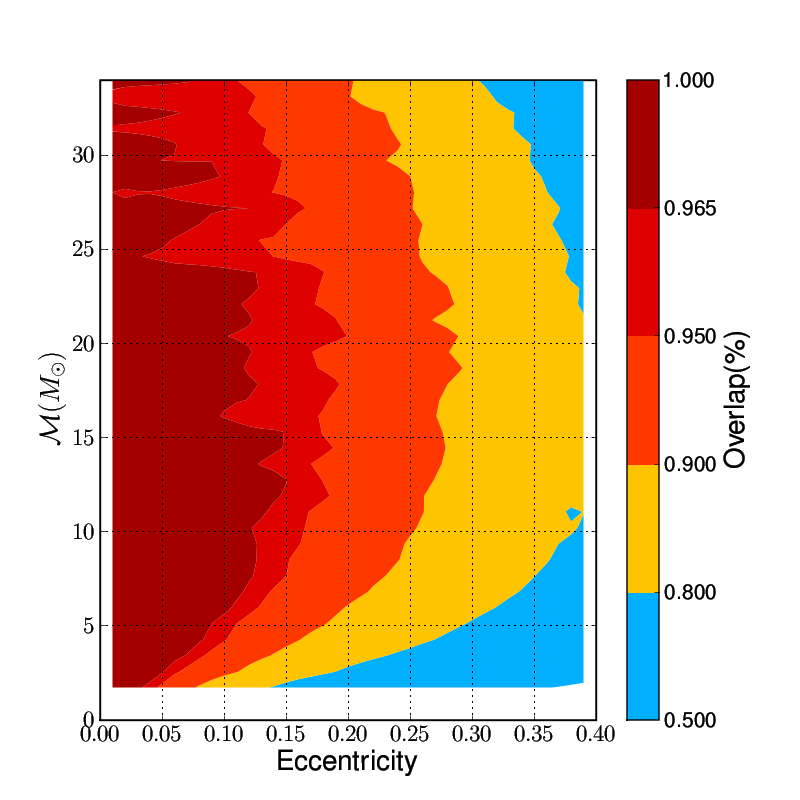}
\caption{\label{fig:chirpmass_match}Matches between the simulated eccentric waveform and a template bank made of non-eccentric waveforms. The eccentricity is uniformly distributed in the range [0,0.4]. The y-axis represent the chirpmass parameter. The data used are identical to those of Fig.~\ref{fig:ligoI}, \ref{fig:ligoa}, \ref{fig:ego} and \ref{fig:virgo}, where the design sensitivity curves are from top left to bottom right : initial LIGO, advanced LIGO, Einstein Telescope and VIRGO.}
\end{figure*}

\begin{figure*}[tbh]
\centering
\includegraphics[width=3in,height=1.8in]{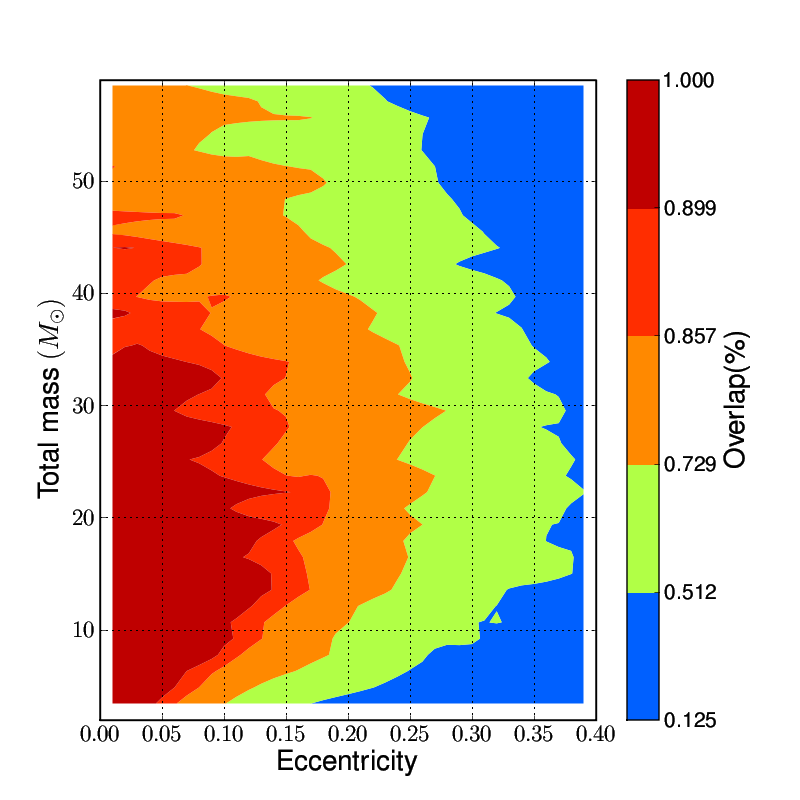}
\includegraphics[width=3in,height=1.8in]{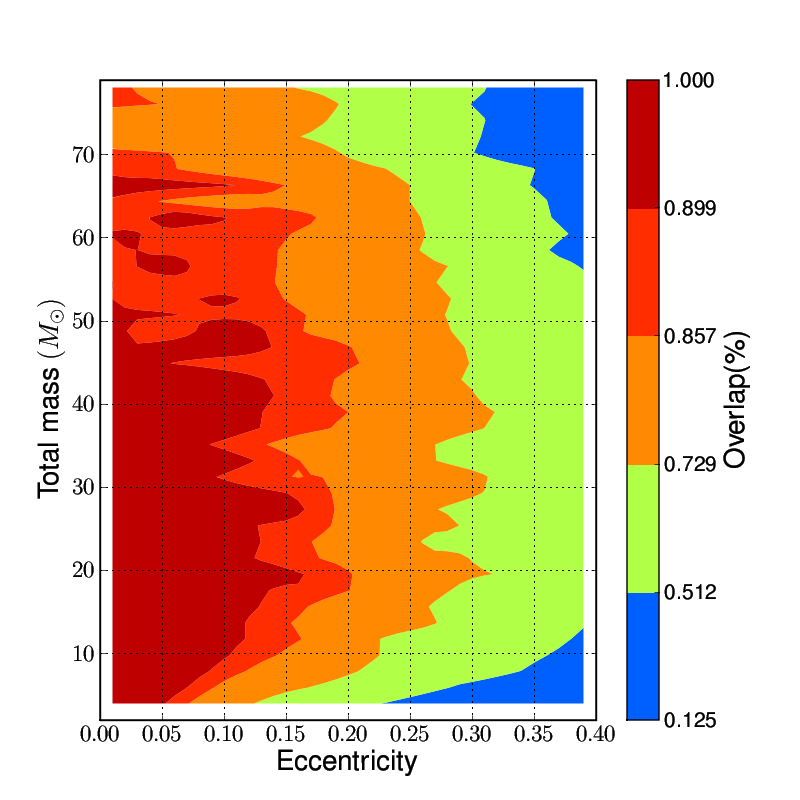}\\
\includegraphics[width=3in,height=1.8in]{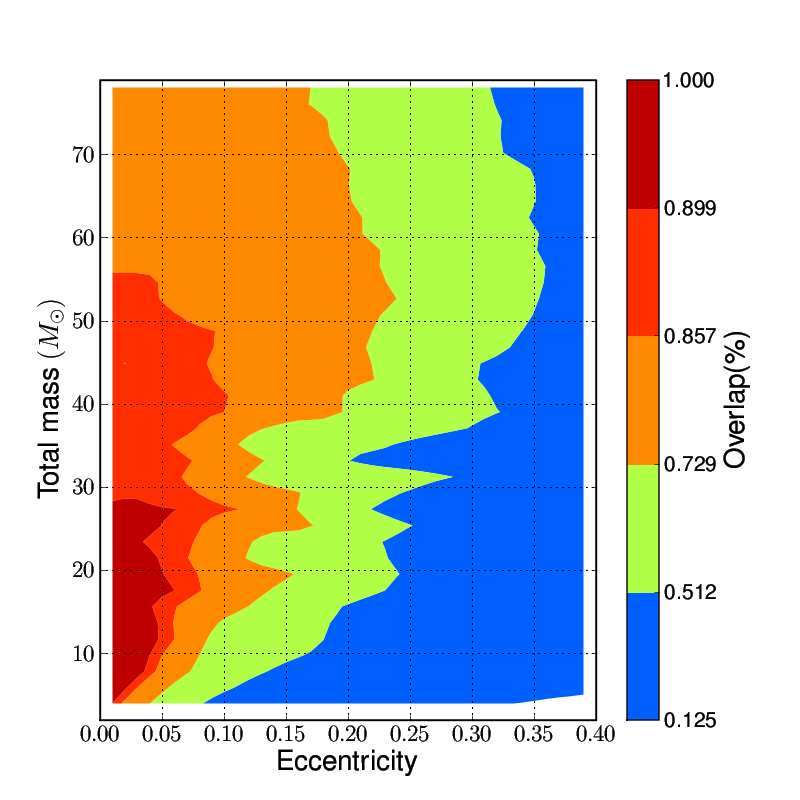}
\includegraphics[width=3in,height=1.8in]{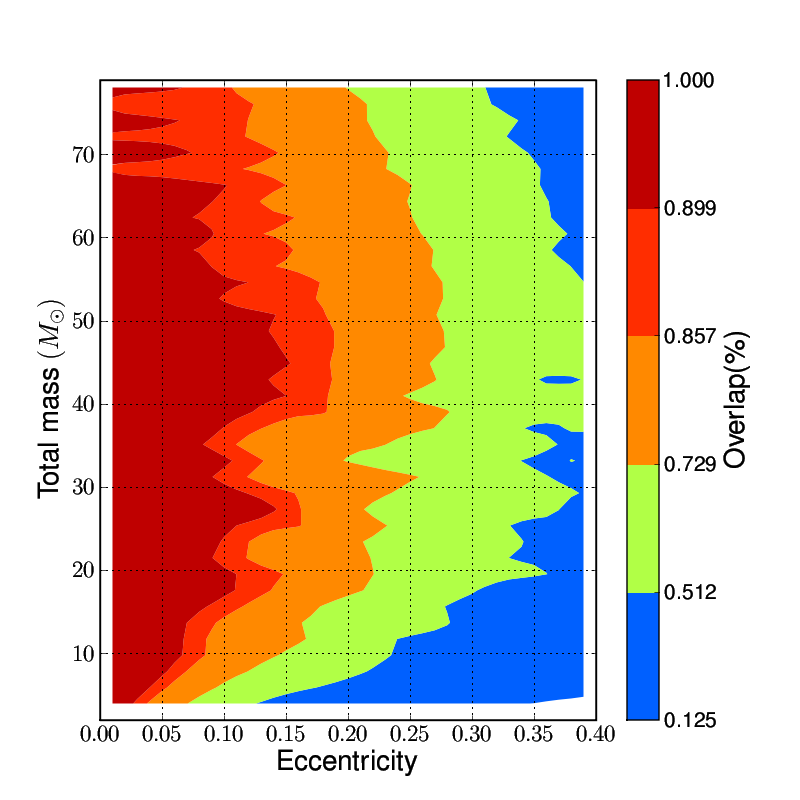}
\caption{\label{fig:lossdetection}Loss of detection. Simulated waveform have eccentricity in the range [0, 0.4]. Template bank is made of non-eccentric waveforms. From top left to bottom right, we used initial LIGO, advanced LIGO, Einstein Telescope and VIRGO design sensitivity curves. The data used are identical to those of Fig.~\ref{fig:chirpmass_match}}
\end{figure*}

\begin{figure*}[tbh]
\centering
\includegraphics[width=3in,height=1.8in]{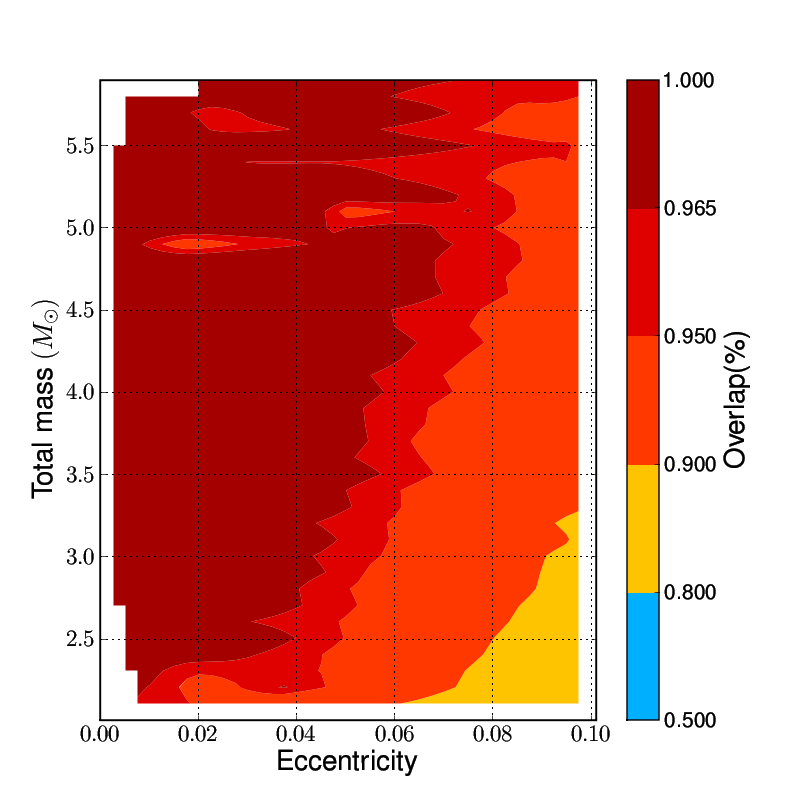}
\includegraphics[width=3in,height=1.8in]{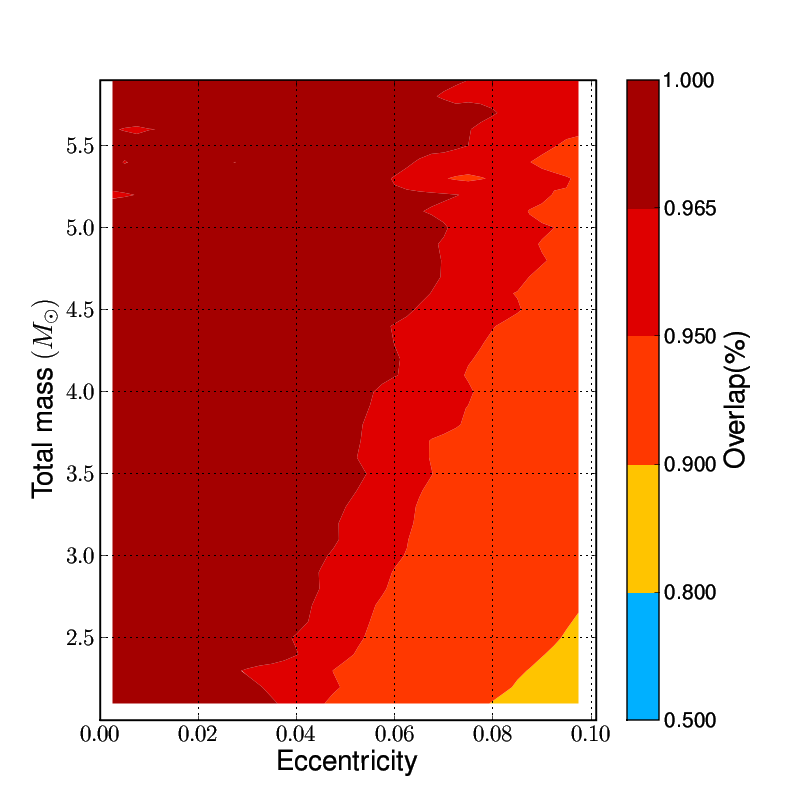}
\\
\includegraphics[width=3in,height=1.8in]{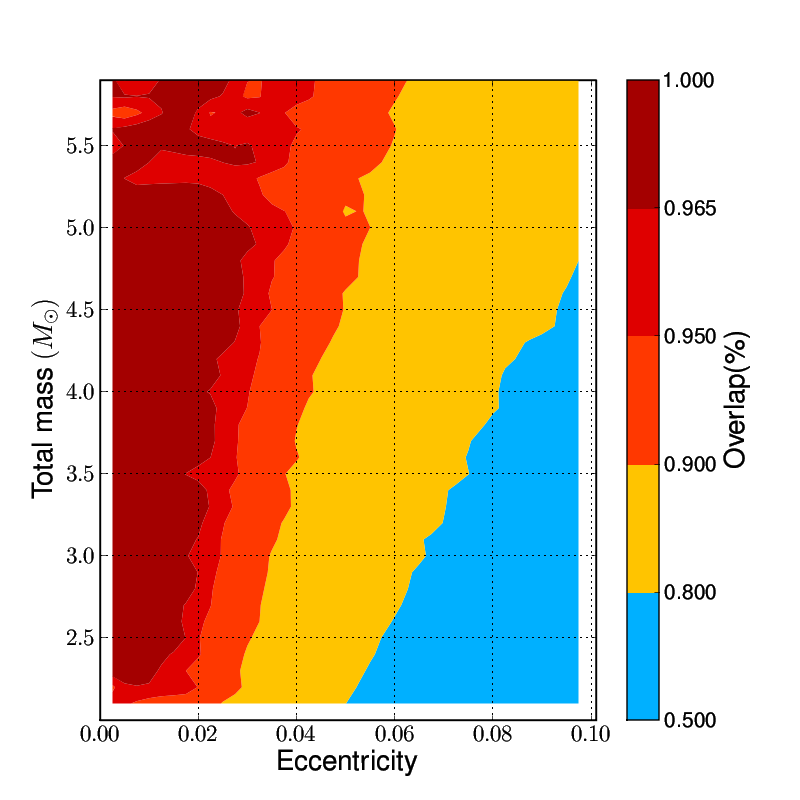}
\includegraphics[width=3in,height=1.8in]{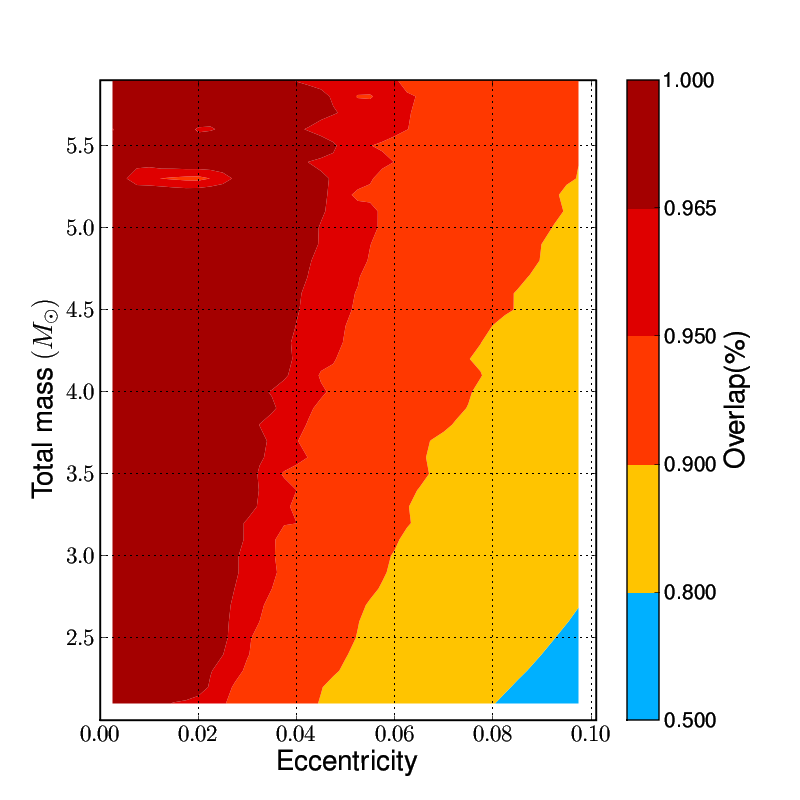}
\caption{\label{fig:bns} Matches between the simulated eccentric waveform (BNS) and a template bank made of non-eccentric waveforms. The initial eccentricity is uniformly distributed in the range [0, 0.1]. From top left to bottom right, we used initial LIGO, advanced LIGO, Einstein Telescope and VIRGO design sensitivity curves.}
\end{figure*}

In Fig.~\ref{fig:chirpmass_match} we provide the results shown in Sec.\ref{sec:simulations} in term of chirp mass rather than the total mass. See the text for a full description of the simulation parameters. We also transform the results into loss of detection in Fig.~\ref{fig:lossdetection}.
Finally, we further explore BNS region by carrying out a simulation restricted to the mass range $M \in [2,6]M_\odot$, and the results of which are shown in Fig.\ref{fig:bns}.

\label{theend}
\end{document}